\documentclass[aps,prl,floatfix,twocolumn,superscriptaddress,nofootinbib]{revtex4-1}
\usepackage{amsmath}
\usepackage{latexsym}
\usepackage{graphicx}
\usepackage{color}
\usepackage{multirow}
\usepackage{array}
\usepackage{rotating}
\usepackage{bbold}
\usepackage{setspace}
\usepackage{colortbl}
\usepackage{endnotes}

\makeatletter
    \def\CT@@do@color{%
      \global\let\CT@do@color\relax
            \@tempdima\wd\z@
            \advance\@tempdima\@tempdimb
            \advance\@tempdima\@tempdimc
    \advance\@tempdimb\tabcolsep
    \advance\@tempdimc\tabcolsep
    \advance\@tempdima2\tabcolsep
            \kern-\@tempdimb
            \leaders\vrule
                    \hskip\@tempdima\@plus  1fill
            \kern-\@tempdimc
            \hskip-\wd\z@ \@plus -1fill }
    \makeatother
    \definecolor{lightgray}{gray}{0.7}
    \newcommand{\bgd}{\cellcolor{lightgray}}
\newcommand*\rfrac[2]{{}^{#1}\!/_{#2}}
\newcommand{\dee}{\mathrm{d}}

\newcommand{\ket}[1]{|#1\rangle}

\newcommand{\und}[1]{$_\mathrm{#1}$}
\begin{document}

\title{Ultrafast time-division demultiplexing of polarization-entangled photons}
\author{John M. Donohue}
\email{jdonohue@uwaterloo.ca}
\address{Institute for Quantum Computing and Department of Physics \&
Astronomy, University of Waterloo, Waterloo, Canada, N2L 3G1}
\author{Jonathan Lavoie}
\address{Institute for Quantum Computing and Department of Physics \&
Astronomy, University of Waterloo, Waterloo, Canada, N2L 3G1}
\address{Group of Applied Physics, University of Geneva, CH-1211 Gen\`{e}ve 4, Switzerland}
\author{Kevin J. Resch}
%\email{kresch@iqc.ca}
\address{Institute for Quantum Computing and Department of Physics \&
Astronomy, University of Waterloo, Waterloo, Canada, N2L 3G1}

\begin{abstract}
Maximizing the information transmission rate through quantum channels is essential for practical implementation of quantum communication. Time-division multiplexing is an approach for which the ultimate rate requires the ability to manipulate and detect single photons on ultrafast timescales while preserving their quantum correlations. Here we demonstrate the demultiplexing of a train of pulsed single photons using time-to-frequency conversion while preserving their polarization entanglement with a partner photon. Our technique converts a pulse train with 2.69~ps spacing to a frequency comb with 307~GHz spacing which may be resolved using diffraction techniques. Our work enables ultrafast multiplexing of quantum information with commercially available single-photon detectors.
\end{abstract}

\maketitle

Quantum communication promises unconditionally secure information transmission by exploiting fundamental features of quantum mechanics~\cite{bennett1984quantum}.  For many protocols, transmission channels capable of distributing entanglement between distant parties are required~\cite{bennett1993teleporting,ekert1991quantum,acin2006bell}. Furthermore, to be practical, these protocols must allow communication at high rates.  One strategy which has successfully increased transmission rates in classical telecommunication is multiplexing, where ancillary degrees of freedom are utilized to carry independent modes co-propagating through a single physical link, such as an optical fibre~\cite{brackett1990dense,kawanishi1998ultrahigh}. Some of these techniques have been adapted to quantum scenarios ~\cite{townsend1997simultaneous,brassard2003multiuser,chen2009stable,qi2010feasibility,choi2010quantum,sasaki2011field,herbauts2013demonstration} and lay the groundwork for future quantum communication networks. \let\thefootnote\relax\footnote{This work also appears as \emph{Phys. Rev. Lett.} \textbf{113}, 163602 (2014).}

Time-division multiplexing~\cite{kawanishi1998ultrahigh} uses the arrival time of light pulses relative to an external clock to distinguish multiple communication modes. It is compatible with fibre-optic systems and is robust against birefringent effects. The delay between subsequent pulses must be greater than the timing jitter of the detection system to avoid cross-talk between signals; for high rates, the delay must also be greater than detector dead time to detect photons from subsequent pulses. State-of-the-art single photon counting detectors have demonstrated 30~ps timing jitter and nanosecond-scale dead times~\cite{hadfield2009single}. However, it is possible in principle to distinguish between two pulses as long as they are separated by their coherence time, which can be orders of magnitude smaller in ultrafast applications. Single-photon measurement techniques for these timescales are therefore critical to optimize the quantum information capacity.

Techniques incorporating short laser pulses and nonlinear optical effects are key to manipulating light on ultrafast timescales~\cite{bennett1999upconversion,walmsley2009characterization,foster2009ultrafast,shayovitz2012high}.  In the quantum regime, such methods have enabled single- and entangled-photon frequency conversion~\cite{kim2001quantum,vandevender2004high,langrock2005highly,tanzilli2005photonic,ramelow2012polarization}, all-optical routing of quantum information~\cite{vandevender2007high,hall2011ultrafast}, and ultrafast coincidence measurement for biphotons~\cite{pe2005temporal,harris2007chirp,o2009time}.  Additionally, ultrafast pulse shaping provides a diverse set of tools to tailor nonlinear optical interactions for customizing quantum optical waveforms~\cite{kielpinski,eckstein2011quantum,rakher2011simultaneous,lavoie13comp,lukens2014orthogonal}, having found application in realizing coherent time-bin measurements on the picosecond timescale~\cite{donohue13}.

\begin{figure}[b!]
  \begin{center}
       \includegraphics[width=1.0\columnwidth]{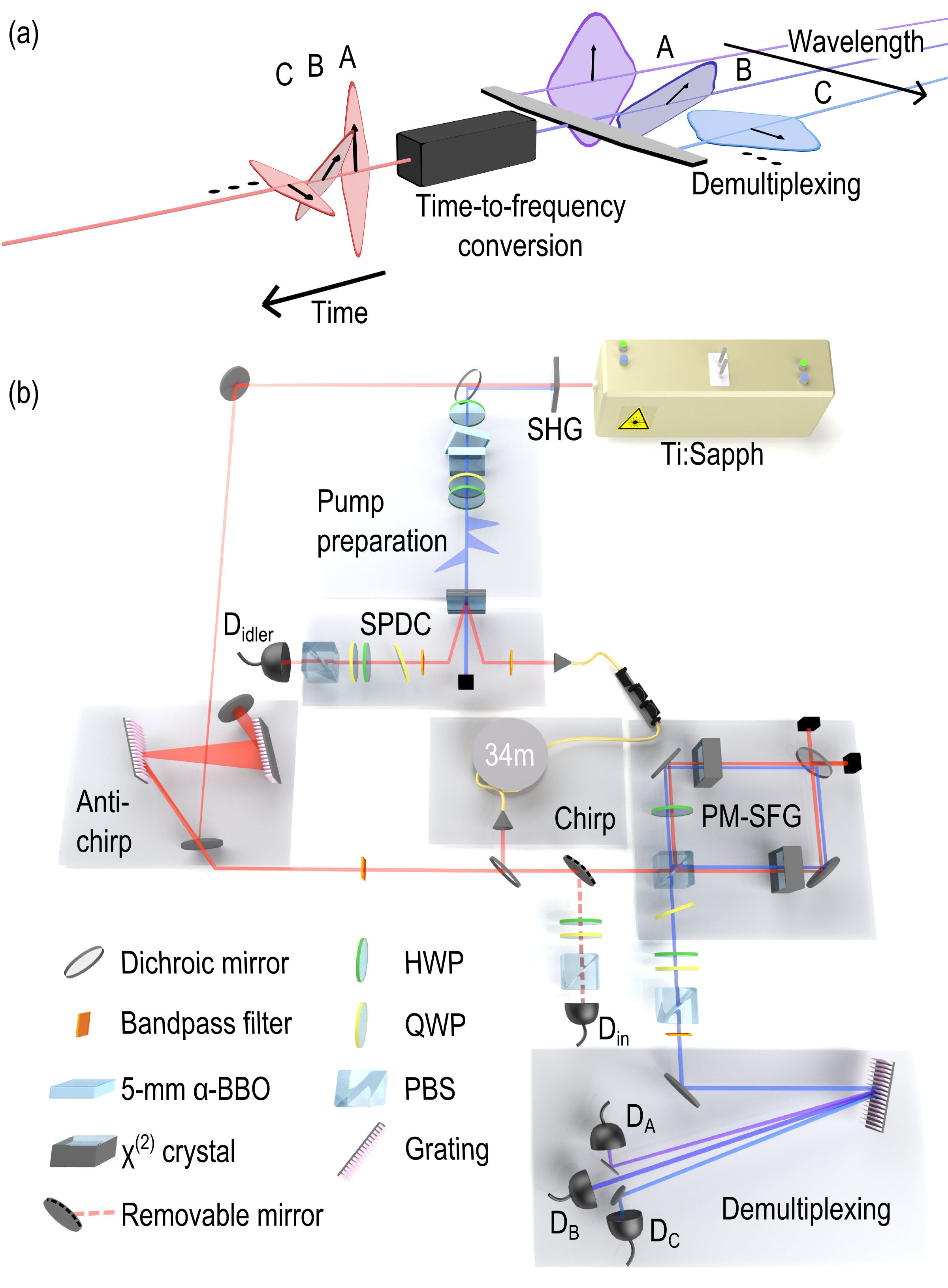}
  \end{center}
 \caption{\textbf{Time-to-frequency conversion concept and experimental setup.} (a) A train of temporally narrow polarized photonic signals A-C are converted into a comb of spectrally narrow and correspondingly polarized photons with a central frequency dependent on their time of arrival. The different frequency modes may then be demultiplexed using diffraction techniques. (b) Two $\alpha$-BBO crystals and a series of wave plates prepared a train of pump pulses $2.69$~ps apart, which were then used to create a pulse sequence of polarization-entangled states through SPDC. The single photons were chirped in single-mode fibre and combined with an anti-chirped strong escort pulse using a dichroic mirror. This beam was then focused in two 10-mm BiBO crystals arranged in a Sagnac configuration for polarization-maintaining sum-frequency generation (PM-SFG). The polarizations of the output photons were measured, and the three signals were then separated with a diffraction grating and coupled to detectors D\und{A-C}. A removable mirror to D\und{in} enabled measurement of the input state.}\label{setup}
\end{figure}

Drawing from these techniques, here we show a method for demultiplexing a rapidly pulsed sequence of polarization-encoded quantum states (Fig.~\ref{setup}a).  Any attempt to directly measure the polarization state of an individual pulse with a photon counter will be subject to crosstalk from the other pulses due to the limited detector time resolution, appearing as an incoherent mixture of the different states. We employ polarization-maintaining sum-frequency generation (SFG) with chirped pulses as a time-to-frequency converter to map ultrafast-scale time delays to measurable frequency shifts, thus allowing the individual quantum states to be read out using conventional diffraction techniques and photon detectors. Furthermore, our method manipulates the time-frequency characteristics of polarization-entangled photons, compressing their spectral bandwidth while preserving entanglement.

Our approach is based on sum-frequency generation between a chirped single photon and an oppositely chirped (anti-chirped) escort laser pulse.
The spectrum of the SFG signal for strongly chirped pulses is much narrower than that of the input light and the frequency produced is linearly dependent on the relative delay between the pulses~\cite{BC_oppositechirp_1,BC_oppositechirp_2,kaltenbaek2008quantum,lavoie13comp}.
We quantify the dispersion applied using the chirp parameter, $A$, defined as $A=\frac{1}{2}\frac{\dee^2 \phi}{\dee\omega^2}$ where $\phi(\omega)$ is the spectral phase.  If the chirps applied are equal and opposite, then the RMS spectral bandwidth of the SFG signal is $\sigma_{\mathrm{SFG}} \le 1/(2 \sqrt{2} A \sigma)$, where $\sigma$ is the smaller of the input bandwidths, and the frequency shift is $\Delta\omega=\tau/(2A)$ away from the sum of the input centre frequencies.  Through this mechanism, chirped-pulse upconversion maps a train of temporally separated pulses into a comb of distinct frequencies.

We require that entanglement is preserved through this time-to-frequency conversion process.  Because of phase-matching considerations, sum-frequency generation in nonlinear crystals is typically efficient for only a specific set of input polarizations, effectively measuring the polarization and destroying coherence.  Preservation of the polarization, and hence the entanglement, can be achieved using a coherent superposition of two sum-frequency processes \cite{ramelow2012polarization}. A full theoretical description of chirped-pulse upconversion applied to a train of polarization-entangled states can be found in the supplemental material.

We create photon pairs using spontaneous parametric down-conversion (SPDC, Fig.~\ref{setup}b).  The pump is produced through second-harmonic generation (SHG) of an 80~MHz titanium-sapphire (Ti:Sapph) femtosecond laser and has a centre wavelength of 394.7~nm with a 1.45~nm full-width at half-maximum bandwidth (FWHM).  Down-conversion is produced in a pair of orthogonally oriented 1-mm $\beta$-barium borate (BBO) crystals cut for type-I down-conversion~\cite{kwiat99spdc,kim2000high}. The source converts pump photons in the polarization state ${\alpha\ket{H}+\beta\ket{V}}$ into down-converted pairs in the polarization state ${\beta\ket{HH}+\alpha\ket{VV}}$, where $\alpha$ and $\beta$ are complex numbers; this can be a separable or entangled state depending on the polarization of the pump.

To create a dense train of pulsed photon pairs, we pass the pump through a series of rotatable birefringent crystals (Pump preparation, Fig.~\ref{setup}b). As the pump propagates through each crystal, the component polarized along the fast axis will lead the one polarized along the slow axis. If the temporal walkoff between these components is greater than the coherence time of the pump, the pump will exit as two pulses which are distinguishable in arrival time relative to a reference from the ultrafast laser source. Using $n$ crystals of identical birefringence, a train of $n+1$ pulses may be created; if the crystal lengths differ, it is possible to create up to $2^n$ pulses~\cite{dromey2007generation}. This prepared pump creates a train of pulsed down-conversion, where the polarization state of each pair is determined by the polarization of the corresponding pump pulse.  To create up to three temporally distinct down-conversion signals, labelled A-C from earliest to latest, we use two 5-mm $\alpha$-BBO crystals cut for maximum birefringence; each apply a relative time delay of $(2.69\pm0.17)$~ps between orthogonal polarization modes. A complete description of the pump preparation setup and down-conversion scheme may be found in the supplemental material.

The signal photons pass through an interference filter centred at 809.06~nm with a 3.9~nm (or 1.8~THz) bandwidth (FWHM) before coupling into 34~m of single-mode fibre, applying positive dispersion corresponding to a chirp parameter of $A=(696\pm3)\times10^3$~fs$^2$. Using a grating-based compressor~\cite{treacy1969optical}, matched negative dispersion is applied to a 225~mW escort pulse at 786.2~nm with a 6.3~nm bandwidth (FWHM). The signal photons and escort pulse are then combined into a single beam with a dichroic mirror.

In order to implement polarization-maintaining SFG, we use a Sagnac-type interferometer (PM-SFG, Fig.~\ref{setup}b). In this configuration, the horizontally and vertically polarized components of the signal photon are split on a polarizing beamsplitter (PBS) and the vertical component is rotated to horizontal polarization using an achromatic half-wave plate. Each beam is then upconverted independently in 10-mm of bismuth borate (BiBO) cut at $150.9^\circ$ for type-I SFG. The SFG signal continues inside the Sagnac loop while the remaining escort is removed using a dichroic mirror. The horizontal component is flipped on the same achromatic half-wave plate and the two components are coherently recombined on the input PBS. A tilted quarter-wave plate sets the phase of the upconverted signal, ensuring that coherent superpositions of $\ket{H}$ and $\ket{V}$ are also maintained.  The internal SFG efficiency was estimated to be 0.3\%. The Sagnac geometry enables passive phase stability, preserving the input polarization state through the sum-frequency process over the 32-hour runtime of the experiment.

\begin{figure}[b!]
  \begin{center}
         \includegraphics[width=1.0\columnwidth]{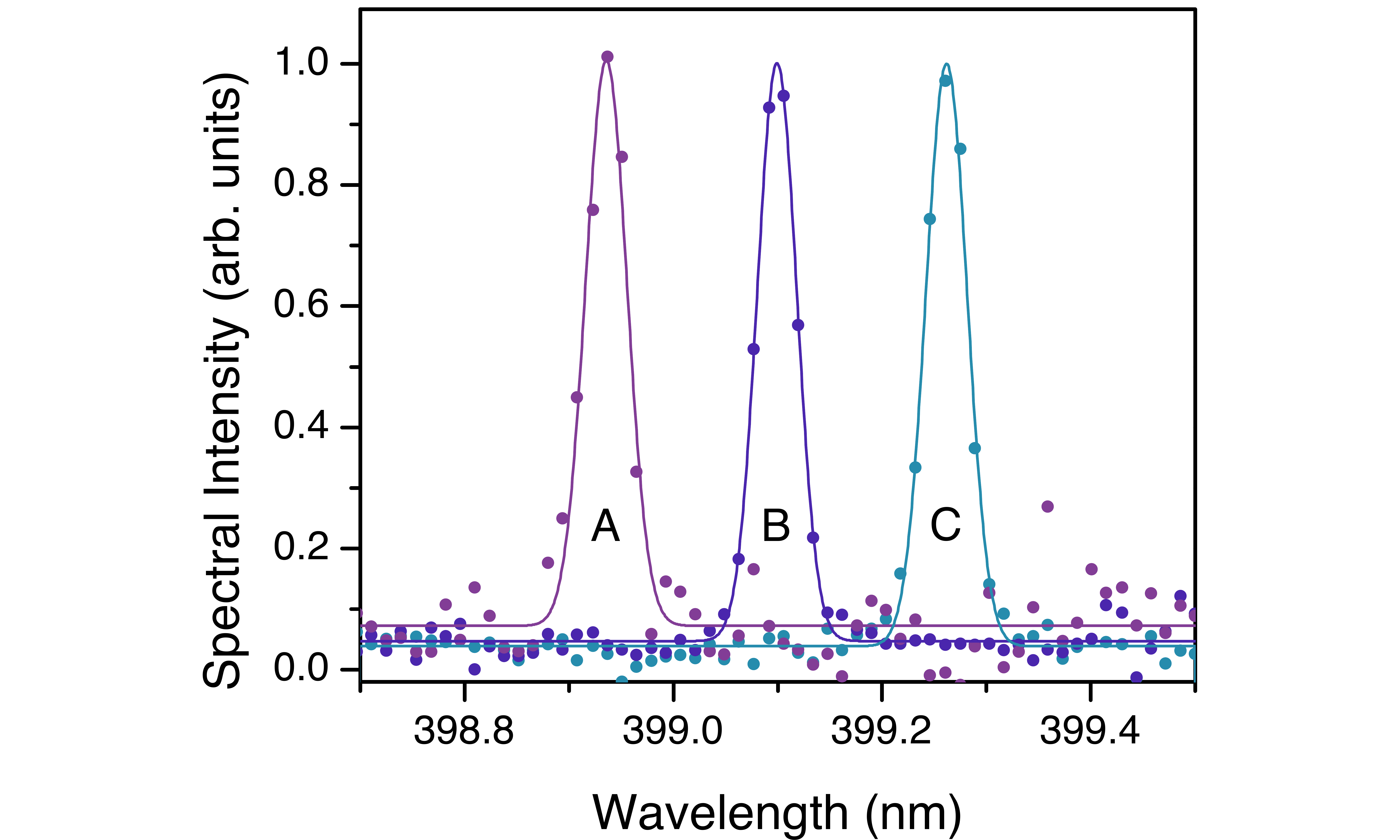}
  \end{center}
 \caption{\textbf{Upconverted single-photon spectra for each temporal mode.} We prepared the pump to maximize the count rate in each of the three temporal modes and measured the spectra shown (with background subtraction). The time delay between the modes maps each to a distinct central wavelength and the spectral bandwidth is compressed by a factor of 20 relative to the input.}\label{spectra}
\end{figure}

After polarization measurement, the remaining near-infrared and escort second harmonic were removed with a bandpass filter. The signals were then separated with a 3600-lines/mm diffraction grating in near-Littrow configuration and allowed to propagate for 4.3~m in free space before being coupled via multimode fibre into three separate detectors, D$_\mathrm{A-C}$. The combined diffraction and coupling efficiency was measured to be approximately 13\%. The measured single-photon spectra were found to have an average bandwidth of $(0.047\pm0.007)$~nm, or equivalently $(88\pm13)$~GHz (Fig.~\ref{spectra}). The spectra measured in modes A-C had respective central wavelengths of $398.936$~nm, $399.099$~nm, and $399.262$~nm. This clearly shows that the three down-conversion pulses, $2.69$~ps apart, were mapped to three distinct wavelengths separated by $(0.163\pm0.007)$~nm, or equivalently $307$~GHz. This spacing is on the same order of magnitude as telecommunication standards for dense wavelength-division multiplexing~\cite{recommendation2012g}.

\begin{figure}[b]
  \begin{center}
       \includegraphics[width=1.0\columnwidth]{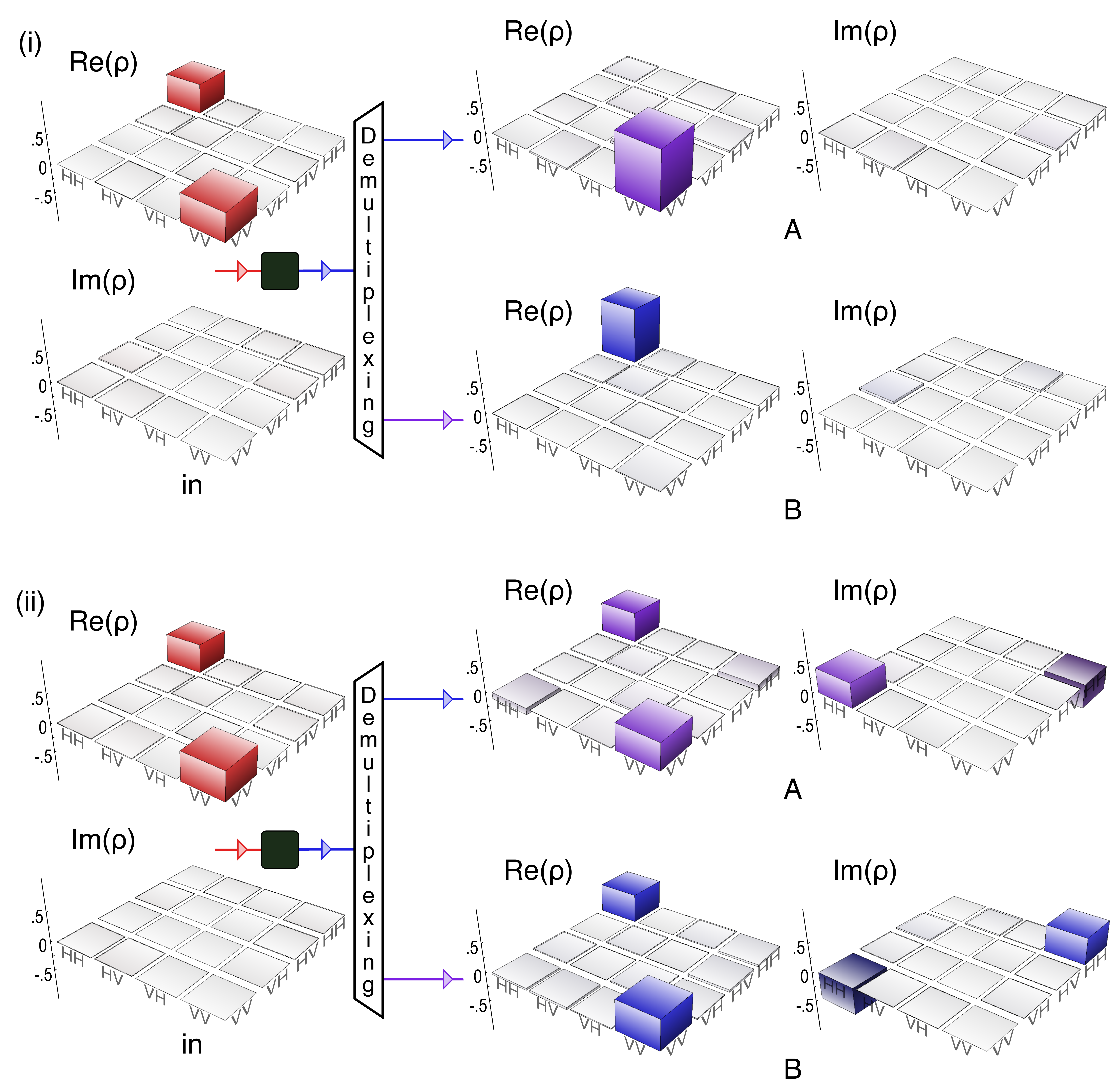}
  \end{center}
 \caption{\textbf{Demultiplexing two orthogonal states.} With the pump prepared in modes A and B to either produce (i) orthogonal separable states or (ii) orthogonal maximally entangled states, the density matrices measured before time-to-frequency conversion (left) appear the same, with negligible coherences. After demultiplexing, the experimentally reconstructed output density matrices (right) are revealed to describe vastly different quantum states, which are separable in case (i) but show a high degree of entanglement in case (ii).}\label{revelation}
\end{figure}

To characterize the preservation of entanglement through our setup, we first prepared the pump to produce the maximally entangled state ${\ket{\Phi^+}=\frac{1}{\sqrt{2}}\left(\ket{HH}+\ket{VV}\right)}$ in a \emph{single} temporal mode at a time. We performed two-photon polarization state tomography~\cite{james2001measurement} both before and after upconversion using an overcomplete set of 36 projective measurements, corresponding to the polarization states $\ket{H}$, $\ket{V}$, ${\ket{\pm}=\frac{1}{\sqrt{2}}\left(\ket{H}\pm\ket{V}\right)}$, and ${\ket{{\pm}i}=\frac{1}{\sqrt{2}}\left(\ket{H}\pm i\ket{V}\right)}$. Because of the polarization-dependent diffraction efficiency of our grating, we performed projective polarization measurements before diffraction. A removable mirror was used to couple the single-photon signal into D$_\mathrm{in}$ to characterize the input state, which was found to have an average fidelity~\cite{jozsa1994fidelity} of 96.2\% with $\ket{\Phi^+}$ over the three potential modes and an average tangle~\cite{coffman2000distributed} of $0.88$. The upconverted states were reconstructed without background subtraction and found to have fidelities $(88.6\pm0.3)$\%, $(95.1\pm0.3)$\%, and $(92.9\pm0.4)$\% with $\ket{\Phi^+}$ and tangles of $0.737\pm0.020$, $0.828\pm0.011$, and $0.836\pm0.015$, for modes A-C respectively, where the error bars are determined by Monte Carlo simulation assuming Poissonian counting statistics. These two figures of merit explicitly demonstrate that quantum correlations are maintained through the bandwidth compression process.

We next prepared the pump to produce down-converted states in modes A and B. We studied the case (i) where the pump was set to produce the separable states $\ket{VV}$ and $\ket{HH}$ (Fig.~\ref{revelation}i), and the case (ii) where the pump was set to produce the maximally entangled states ${\ket{\Phi^{+i}}=\frac{1}{\sqrt{2}}\left(\ket{HH}+i\ket{VV}\right)}$ and ${\ket{\Phi^{-i}}=\frac{1}{\sqrt{2}}\left(\ket{HH}-i\ket{VV}\right)}$ (Fig.~\ref{revelation}ii), in modes A and B respectively.  The reconstruction from the coincidence measurements between D\und{in} and D\und{idler} produced the density matrix on the left-hand side of Fig.~\ref{revelation}, with large populations in $\ket{HH}$ and $\ket{VV}$ but negligible coherence; both reconstructions have fidelities of 98\% with an equal mixture of $\ket{HH}$ and $\ket{VV}$. This arises because the detector is not fast enough to resolve the pulses, instead measuring a mixture of the two signals and obfuscating the underlying quantum coherences of the individual states.  By measuring the photons after the upconversion setup, the density matrices shown on the right side of Fig.~\ref{revelation} were reconstructed.  The density matrices in case (ii) exhibit large quantum coherences, which are required for entanglement, while those for case (i) do not, as expected for separable states.  Indeed, the density matrices reconstructed in case (i) have an average fidelity of ($93.6\pm0.3$)\% with the target separable states, and those in case (ii) have an average fidelity of ($91.2\pm0.5$)\% with the expected maximally entangled states and an average tangle of $0.714\pm0.014$.

\begin{figure}[b]
  \begin{center}
           \includegraphics[width=1.0\columnwidth]{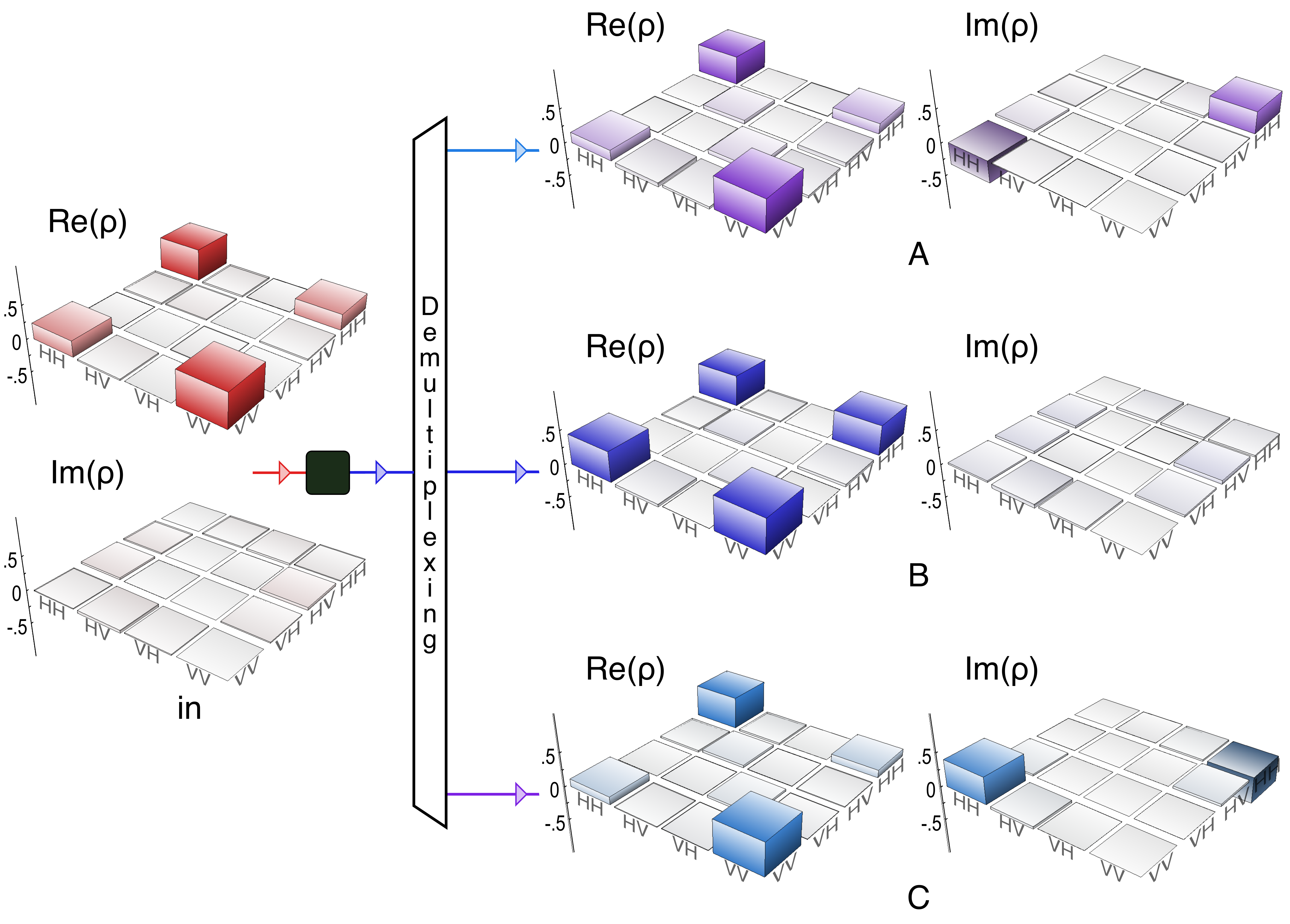}
  \end{center}
 \caption{\textbf{Demultiplexing three entangled states.} The pump was prepared to produce a train of three maximally entangled states. Weak coherences are seen in the density matrix measured before time-to-frequency conversion (left), with a calculated tangle of $0.21$. After being demultiplexed, all three experimentally reconstructed density matrices show much stronger coherence and larger entanglement, with tangles of $0.40$, $0.72$, and $0.58$ in modes A-C respectively.}\label{distribution}
\end{figure}

We then prepared the pump to produce maximally entangled states into all three modes, using the pump polarization sequence $\ket{{-}i}$, $\ket{+}$, and $\ket{{+}i}$ for modes A-C.  We measured the states initially and after the upconversion process, shown in (Fig.~\ref{distribution}).  The initial state has fidelity $97.6\%$ with the non-maximally entangled mixed state resulting from an incoherent mixture of the three expected maximally entangled states in modes A-C with weighting 0.25, 0.5, and 0.25, determined by the ratios of the intensities of the three pump pulses.  The output states each exhibit different quantum correlations yet are all highly entangled, with fidelities of $(77.3\pm0.9)\%$, $(91.5\pm0.4)\%$, and $(86.1\pm0.7)\%$ with the expected maximally entangled states and tangles of $(0.40\pm0.2)$, $(0.720\pm0.013)$, and $(0.58\pm0.02)$  for modes A-C, respectively.  The coincidence rates for modes A and C were half that of mode B due to the distribution of pump power, and their reconstructed states were thus more affected by background noise; however, crosstalk between signals was not a significant issue in our experiment.  Additional experimental results for different pump preparations may be found in the supplemental material.

We have demonstrated the conversion of a train of up to three temporally spaced single-photon pulses to a comb of distinct frequencies while maintaining quantum correlations in polarization.  We have shown that this method can distinguish picosecond-separated single photons using detectors with nanosecond-scale time resolution.  Improvements to the efficiency may be possible through the use of periodically poled nonlinear materials~\cite{vandevender2004high,langrock2005highly} and cavity enhancements~\cite{sensarn2009resonant}.  With higher conversion efficiencies, this ultrafast readout of time-division-multiplexed entangled quantum signals could be used to increase the density of quantum information carried through a single physical medium or to distribute quantum states throughout a multi-user network by applying time-to-frequency conversion to both signal and idler photons. Our results also demonstrate tunable bandwidth compression of a polarization-entangled photon~\cite{lavoie13comp}.  More generally, our work demonstrates how shaped laser pulses may be used to manipulate the spatiotemporal waveforms of single photons while preserving quantum information.

\begin{acknowledgements}
The authors thank M.~D.~Mazurek, A.~Martin, K.~A.~G.~Fisher, and M.~Agnew for helpful discussions. We are grateful for financial support from the Natural Sciences and Engineering Research Council, Canada Foundation for Innovation, Ontario Centres of Excellence, Industry Canada, the Canada Research Chairs Program, and the Ontario Ministry of Research and Innovation.
\end{acknowledgements}

\newpage
\onecolumngrid

\newpage

%% SUPP MAT HACKS %%
% For section headers starting with S\renewcommand{\thesection}{S.\arabic{section}}
\renewcommand{\thesubsection}{\thesection.\arabic{subsection}}
% Hack for making SOM Equations Conform to Science Format
% e.g. (S1), (S2), etc
% Requires AMS
\makeatletter %% With ams
\def\tagform@#1{\maketag@@@{(S\ignorespaces#1\unskip\@@italiccorr)}}
\makeatother
% Hack for making figures Say \figurename S\thefigure, e.g. Figure S1:
\makeatletter
\makeatletter \renewcommand{\fnum@figure}
{\figurename~S-\thefigure}
\makeatother
\makeatletter
\makeatletter \renewcommand{\fnum@table}
{\tablename~S-\thetable}
\makeatother
% use bibnumfmt to change style at the end of the document
%\renewcommand{\bibnumfmt}[1]{[S#1]}
% citenumfont command adds S to all numbers
%\renewcommand{\citenumfont}[1]{S#1}
\renewcommand{\figurename}{Figure}
%END HACKS
\setcounter{equation}{0}
\setcounter{figure}{0}
\setcounter{table}{0}

\section{Supplemental Material}

\subsection{Theory of single-photon time-to-frequency conversion}\label{TheoreticalDescription}

To understand the demultiplexing process, we consider a strong laser pump propagating in the $\hat{z}$ direction which has been divided into a number of pulses at times $\tau_j$ with well-defined polarizations.  We model the electric field of the pump classically as \begin{equation}\vec{E}_p(t)=\sum_j\int\dee\omega_p\, e^{i\omega_p\tau_j}\left(\cos\theta_j\hat{\epsilon}_H+e^{i\phi_j}\sin\theta_j\hat{\epsilon}_V\right) \sqrt{b_j}E_{0p}\xi(\omega_p)e^{ik_pz-i\omega_pt},
\end{equation} where $E_{0p}$ characterizes the electric field of the initial pump and $\xi(\omega_p)$ its spectrum; $b_j$ represents the pulse amplitudes for the different temporal modes. We describe spontaneous parametric down-conversion in a $\chi^{(2)}$ medium with an undepleted pump using the unitary~\cite{grice1997spectral} \begin{align}\hat{U}_{SPDC}=\exp{\bigg[}i&\gamma_{1}\sum_j\sqrt{b_j}\iiint\dee\omega_s\dee\omega_i\dee\omega_p\delta(\omega_p-\omega_i-\omega_s)\, \nonumber \\ &\left\{e^{i\omega_p\tau_j}\xi(\omega_p)\Phi_{SPDC}(\omega_s,\omega_i,\omega_p)\left[\hat{a}^{\dag(s,H)}_{\omega_s}\hat{a}^{\dag(i,H)}_{\omega_i}\cos\theta_j+\hat{a}^{\dag(s,V)}_{\omega_s}\hat{a}^{\dag(i,V)}_{\omega_i}e^{i\phi_j}\sin\theta_j\right]+\mathrm{h.c.}\right\}{\bigg]},\label{Uspdc}\end{align} where $s$ and $i$ represent the signal and idler modes respectively, $\Phi_{SPDC}$ represents the phasematching function of the medium, and $\gamma_{1}$ is a constant which depends on $E_{0p}$, the nonlinearity, and the length of the medium.  If the signal and idler modes are initially in the vacuum state, this unitary, to first order, will produce photon pairs in the state \begin{equation}\ket{\psi}=\sum_j\sqrt{b_j}\iint\dee\omega_s\dee\omega_i\, e^{i(\omega_s+\omega_i)\tau_j}\xi(\omega_s+\omega_i)\Phi_{SPDC}(\omega_s,\omega_i,\omega_s+\omega_i) \left[\cos\theta_j\ket{H,\omega_s}_s\ket{H,\omega_i}_i+e^{i\phi_j}\sin\theta_j\ket{V,\omega_s}_s\ket{V,\omega_i}_i\right].\end{equation}
%This first-order expansion has assumed that the probability of two down-conversion processes occurring simultaneously is small; indeed, if this were to be a likely occurrence, we would need to also demultiplex the idler photon. However, this regime would also lead to a high likelihood of double-pair emission into a single mode, degrading the quality of the single-photon states.

We model the effect of spectral filters and pulse shaping by applying the functions $f_s(\omega_s)$ and $f_i(\omega_i)$ to the signal and idler amplitudes. We make the assumption that these filter functions are spectrally narrow relative to the pump field and the phasematching function, such that \begin{equation}f_s(\omega_s)f_i(\omega_i)\xi(\omega_s+\omega_i)\Phi_{SPDC}(\omega_s,\omega_i,\omega_s+\omega_i)\approx f_s(\omega_s)f_i(\omega_i),\end{equation} which renders the final state separable in frequency. It is important to work in this regime as energy-time entanglement degrades the effectiveness of bandwidth compression through chirped-pulse sum-frequency generation \cite{lavoie13comp}. Our time-division multiplexed down-converted state may then be written as \begin{equation}\ket{\psi}=\sum_j\sqrt{b_j}\iint\dee\omega_s\dee\omega_i\, e^{i(\omega_s+\omega_i)\tau_j}f_s(\omega_s)f_i(\omega_i) \left[\cos\theta_j\ket{H,\omega_s}_s\ket{H,\omega_i}_i+e^{i\phi_j}\sin\theta_j\ket{V,\omega_s}_s\ket{V,\omega_i}_i\right].\end{equation}

We now subject the signal mode to polarization-maintaining sum-frequency generation, another $\chi^{(2)}$ process which destroys a photon in the signal mode and creates one in the \emph{generated mode} $g$ with the help of an strong \emph{escort} pulse $e$. We model the escort pulse as a strong coherent state and describe this process using the unitary \begin{align}\hat{U}_{SFG}=\exp{\bigg[}i&\gamma_{2}\iint\dee\omega_s\dee\omega_g\, \left\{\alpha(\omega_g-\omega_s)\Phi_{SFG}(\omega_s,\omega_g-\omega_s,\omega_g)\left[\hat{a}^{(s,H)}_{\omega_s}\hat{a}^{\dag(g,H)}_{\omega_g}+\hat{a}^{(s,V)}_{\omega_s}\hat{a}^{\dag(g,V)}_{\omega_g}\right]+\mathrm{h.c.}\right\}{\bigg]},\label{Usfg}\end{align}  where $\alpha(\omega_e)$ is the spectrum of the escort pulse and $\gamma_2$ is a constant describing the strength of the interaction. Assuming broad phasematching in the second crystal (such that $\Phi_{SFG}(\omega_s,\omega_g-\omega_s,\omega_g)\approx1$), we expand the state to first order in $\gamma_2$ and post-select on the successful generation of a sum-frequency field, \begin{equation}\ket{\psi_f}=\sum_j\sqrt{b_j}\iint\dee\omega_i\dee\omega_g\, e^{i\omega_i\tau_j}f_i(\omega_i)f_{g,j}(\omega_g) \left[\cos\theta_j\ket{H,\omega_i}_i\ket{H,\omega_g}_g+e^{i\phi_j}\sin\theta_j\ket{V,\omega_i}_i\ket{V,\omega_g}_g\right],\end{equation} where the upconverted state maintains the initial polarization correlations of the down-conversion and the spectrum of the sum-frequency photon \emph{in each individual mode} is given by \begin{equation}f_{g,j}(\omega_g)=\int\dee\omega_se^{i\omega_s\tau_j}f_s(\omega_s)\alpha(\omega_g-\omega_s).\label{SFGspecDef}\end{equation}

If the signal photon and the strong escort pulse both have Gaussian spectra and are oppositely chirped, their spectral amplitudes may be described (ignoring phase) as \begin{align}f_s(\omega_s)&=\frac{1}{(2\pi\sigma_s^2)^\frac{1}{4}}e^{-\frac{(\omega_s-\omega_{0s})^2}{4\sigma_s^2}}e^{iA(\omega_s-\omega_{0s})^2}\\ \alpha(\omega_e)&=\frac{1}{(2\pi\sigma_e^2)^\frac{1}{4}}e^{-\frac{(\omega_e-\omega_{0e})^2}{4\sigma_e^2}}e^{-iA(\omega_e-\omega_{0e})^2},\end{align} and the spectrum of the generated photon may be written as \begin{align}f_{g,j}(\omega_g)&=\sqrt{\frac{2\sigma_e\sigma_s}{\sigma_e^2+\sigma_s^2}} e^{-\frac{1+16A^2\sigma_e^2\sigma_s^2}{4(\sigma_e^2+\sigma_s^2)}\left(\omega_g-\omega_{0g}+\frac{8A\sigma_s^2\sigma_e^2}{1+16A^2\sigma_e^2\sigma_s^2}\tau_j\right)^2} e^{-\frac{\sigma_e^2\sigma_s^2}{(\sigma_e^2+\sigma_s^2)(1+16A^2\sigma_e^2\sigma_s^2)}\tau_j^2} e^{i\left[\frac{A(\sigma_s^2-\sigma_e^2)(\omega_g-\omega_{0g})^2}{\sigma_s^2+\sigma_e^2}+\frac{\tau_j\sigma_s^2(\sigma_s^2+\sigma_e^2)(\omega_g-\omega_{0g})}{\sigma_s^2+\sigma_e^2}\right]}\nonumber\\ &\approx \sqrt{\frac{2\sigma_e\sigma_s}{\sigma_e^2+\sigma_s^2}} e^{-\frac{4A^2\sigma_e^2\sigma_s^2}{\sigma_e^2+\sigma_s^2}\left(\omega_g-\omega_{0g}+\frac{\tau_j}{2A}\right)^2} e^{-\frac{\tau_j^2}{(\sigma_e^2+\sigma_s^2)(16A^2)}} e^{i\left[\frac{A(\sigma_s^2-\sigma_e^2)(\omega_g-\omega_{0g})^2}{\sigma_s^2+\sigma_e^2}+\frac{\tau_j\sigma_s^2(\sigma_s^2+\sigma_e^2)(\omega_g-\omega_{0g})}{\sigma_s^2+\sigma_e^2}\right]},\end{align} where $\omega_{0g}=\omega_{0e}+\omega_{0s}$ and the approximation takes the large chirp limit, $A\sigma_i^2\gg1$. Note that, in this large-chirp approximation, the RMS intensity bandwidth of each spectral component is \begin{equation}\sigma_g\approx\frac{1}{4A}\sqrt{\frac{1}{\sigma_e^2}+\frac{1}{\sigma_s^2}}.\end{equation} Each component is also shifted in central frequency by $\Delta\omega\approx\tau_j/2A$ and has an exponential decay factor proportional to $\tau_j^2$. If we exactly measure a frequency $\omega_{0g}-\tau_m/2A$, the signal found will be proportional to \begin{equation}\left|f_{g,j}\left(\omega_{0g}-\frac{\tau_m}{2A}\right)\right|^2\propto\exp\left[-\frac{2\sigma_e^2\sigma_s^2}{\sigma_e^2+\sigma_s^2}(\tau_j-\tau_m)^2-\frac{\tau_j^2}{8A^2(\sigma_e^2+\sigma_s^2)}\right].\end{equation} To ensure that the crosstalk from neighbouring modes is negligible (i.e. $\left|f_{g,j}\left(\omega_{0g}-\frac{\tau_m}{2A}\right)\right|^2\approx0$ for $j\neq m$), we define $\Delta\tau_{min}$ as the separation between any directly adjacent temporal modes and require that \begin{equation}\Delta\tau_{min}>2\frac{1}{\sqrt{2}}\frac{\sqrt{\sigma_e^2+\sigma_s^2}}{\sigma_e\sigma_s}\label{lowertau}\end{equation} is satisfied, which is equivalent to requiring that the separation is greater than Fourier-limited temporal width of the input pulses. We also define $\Delta\tau_{max}$ as the maximum temporal separation of any individual temporal mode and the escort pulse.  To ensure comparable efficiencies for each temporal mode, we require that the chirped fields overlap well, i.e. \begin{equation}\Delta\tau_{max}<2\sqrt{2}|A|\sqrt{\sigma_e^2+\sigma_s^2}.\label{uppertau}\end{equation}

In our experiment, the separation between modes is $2.69$~ps. Using our experimental parameters, the lower bound of Eq.~S\ref{lowertau} is $0.4$~ps and the upper bound of Eq.~S\ref{uppertau} is $18$~ps. Thus, we are well within the required limits. Again using our experimental parameters, one could, in principle, access up to approximately $\Delta\tau_{max}/\Delta\tau_{min}=45$ distinct modes.

\newpage

\subsection{Details on down-conversion source}\label{DetailsSPDC}

\begin{figure}[h]
\begin{center}
 \includegraphics[width=0.75\columnwidth]{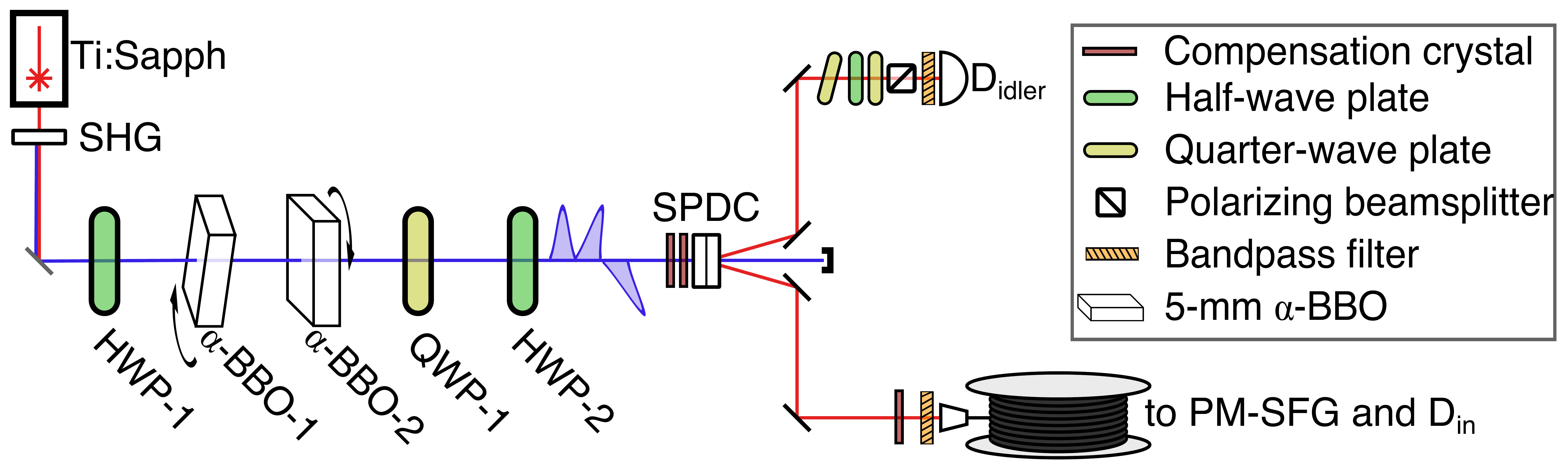}
 \caption{\textbf{Pump preparation and down-conversion.} Schematic of down-conversion setup and pump temporal preparation.}\label{PumpSetup}
 \end{center}
\end{figure}

Our experiment uses a titanium-sapphire laser with a repetition rate of 80~MHz, centre wavelength $790.1\pm0.2$~nm with a bandwidth (FWHM) of $12.27\pm0.08$~nm and an average power of 2.3~W. It was frequency-doubled in 2~mm of bismuth borate (BiBO) cut at 152.4$^\circ$ to a second-harmonic centred at $394.7$~nm with a bandwidth of $1.45\pm0.02$~nm and an average power of 0.6~W. The remaining power of the fundamental Ti:Sapph beam was used as the strong escort pulse, which was filtered to a centre wavelength of $786.2$~nm with a bandwidth of $6.3$~nm to reduce the background after upconversion arising from its second harmonic.

We generated down-conversion in a pair of two orthogonally-oriented 1-mm BBO crystals cut at 30$^\circ$~\cite{kwiat99spdc,lavoie2009experimental} with a full opening angle of approximately 6$^\circ$. In order to compensate for the effects of temporal and spatial walkoff, additional birefringent crystals were required: 1~mm of $\alpha$-BBO and 1~mm of crystal quartz were inserted in the path of the UV pump to correct for temporal walkoff~\cite{kim2000high} and 1~mm of BiBO cut at an angle of 152.6$^\circ$ was inserted in the signal arm to correct for spatial walkoff (Fig.~S-\ref{PumpPrep}). To remove energy-time entanglement, the signal was filtered to $809.06$~nm with a bandwidth of $3.9$~nm and the idler to $770.58$~nm with a bandwidth of $2.27$~nm. A quarter-wave plate at zero degrees was tilted to apply a controllable phase to the idler, aligned such that the state $\ket{\Phi^+}=\frac{1}{\sqrt{2}}\left(\ket{HH}+\ket{VV}\right)$ was detected between D$_\mathrm{idler}$ and the detector before upconversion, D$_\mathrm{in}$.

Near-infrared detectors D\und{idler} and D\und{in} were Perkin-Elmer SPCM-AQ4C photon counting modules, with a quantum efficiency of approximately 50\% near 800~nm and a single-photon timing jitter of approximately 600~ps. Near-UV detectors D\und{A-C} were Hamamatsu H10682-210 photon counting heads, with a quantum efficiency of approximately 30\% around 400~nm and a single-photon timing jitter of approximately 200~ps.

\subsection{Additional experimental results}\label{AdditionalResults}

In the main text, we presented results for six different pump preparations: three of which prepared entangled states in one mode at a time, two prepared orthogonal states in two different modes, and one prepared entangled states in three different modes.  In this section, we present experimental results for two additional three-mode pump preparations as well as details on the settings required to prepare the pump for all cases (Table~S-\ref{PumpPrep}).

The pump passes through two 5-mm $\alpha$-BBO crystals, which each introduce a birefringent delay of $2.69$~ps between pulse components polarized along the fast and slow axes. As the first and last pulse of the train are necessarily polarized on opposite axes in the last crystal, they are necessarily orthogonally polarized. Therefore, if the first pump pulse consists of photons described by the polarization state $\ket{\psi}$, the last pulse must be described by the orthogonal state, $\ket{\psi^\perp}$. In a three-pulse preparation, there are also restrictions on the middle state. To create three pulses with two identical birefringent crystals, two orthogonal pulses must be created in the first crystal. The second crystal will once again split each pulse in two, and the component of the leading pulse on the slow axis of the crystal will overlap in time with the component of the lagging pulse on the fast crystal axis. As these two components were on different axes, they are necessarily orthogonal. In the case where the leading and lagging pulses are of equal amplitude, the middle pulse will be in some polarization state describable as ${\frac{1}{\sqrt{2}}\left(\ket{\psi}+e^{i\phi}\ket{\psi^\perp}\right)}$ and have twice the photon number of the other modes. The parameter $\phi$ may be manipulated by controlling the pump polarization.

\begin{figure}[h]
  \begin{center}
       \includegraphics[width=1\columnwidth]{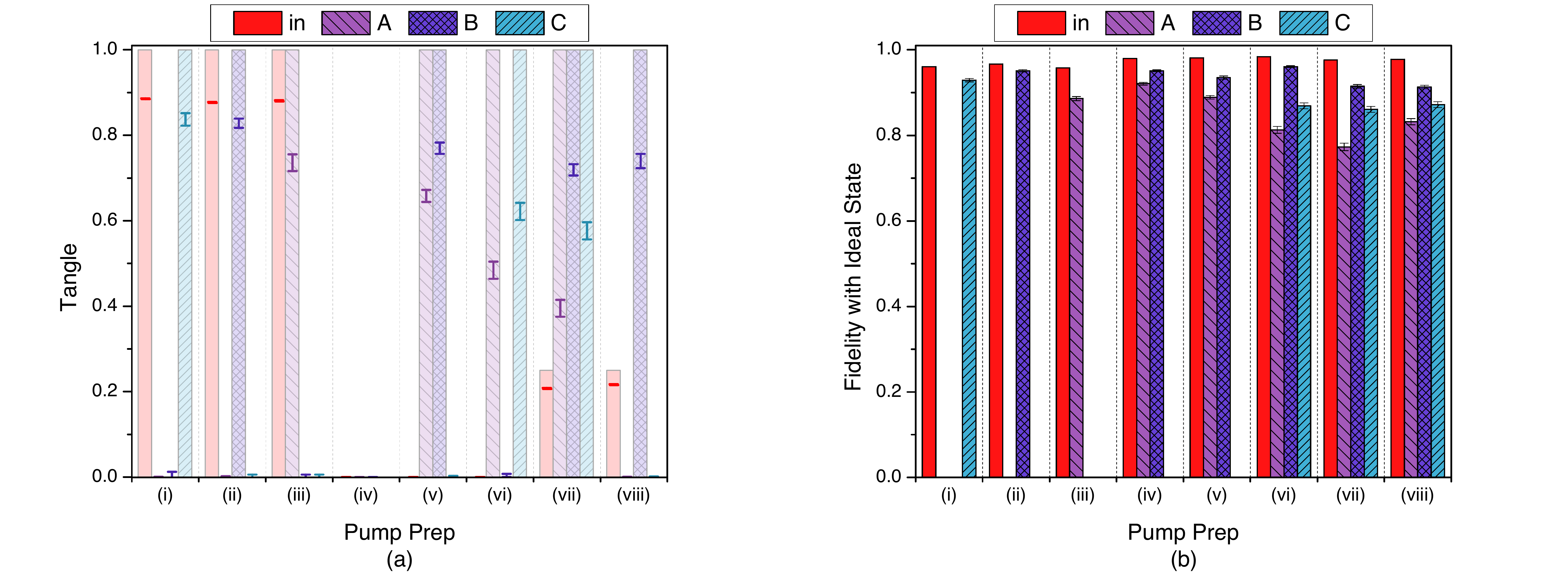}
  \end{center}
 \caption{\textbf{Tangle and fidelity measurements.} (a) Measurements of the tangle~\cite{coffman2000distributed} for each preparation and mode are shown, with their theoretical ideals transparent in the background. Note that the situation where entanglement is expected are consistently many standard deviations (at least twenty) above zero tangle, and that there are cases where no entanglement is measured in the input state yet presents itself in the demultiplexed subsystems. (b) The fidelity~\cite{jozsa1994fidelity} of the reconstructed density matrix with the theoretical ideal is seen to be high for all cases (at minimum 77.3\%). Background modes, i.e. those with no pump pulse, omitted for clarity.}\label{tangle}
\end{figure}

Full measurement results for each of these preparations are displayed in Table~S-\ref{FullResults}, with the tangle and fidelity explicitly plotted in Fig.~S-\ref{tangle}. Each set of tomographic data required 36 projective measurements. Coincidences were recorded for five seconds for the input state. For the single-mode measurements, six loops of thirty-second coincidence measurements were recorded, for a total of three minutes per setting. For all other settings, twelve loops of thirty-second measurements were recorded, for a total of six minutes per setting. The background results presented are the average of two such runs, where the signal was blocked but the idler and escort were unchanged.

\begin{table}[h]
\begin{center}
\begin{tabular}{|c|c|c|c|c|c|c|c|}
\hline
&\multicolumn{3}{c|}{State}&\multicolumn{4}{c|}{Angle}\\
\hline
&A&B&C&$\alpha$-BBO-1&$\alpha$-BBO-2&QWP-1&HWP-2\\
\hline
(i)&-&-&$\ket{\Phi^{+1}}$&$0$&$0$&$\rfrac{\pi}{2}$&$\rfrac{\pi}{8}$\\
(ii)&-&$\ket{\Phi^{+1}}$&-&$\rfrac{\pi}{2}$&$0$&$\rfrac{\pi}{2}$&$\rfrac{\pi}{8}$\\
(ii)&$\ket{\Phi^{+1}}$&-&-&$\rfrac{\pi}{2}$&$\rfrac{\pi}{2}$&$\rfrac{\pi}{2}$&$\rfrac{\pi}{8}$\\
(iv)&$\ket{HH}$&$\ket{VV}$&-&$\rfrac{\pi}{2}$&$\rfrac{\pi}{4}$&$\rfrac{3\pi}{4}$&$\rfrac{\pi}{8}$\\
(v)&$\ket{\Phi^{-i}}$&$\ket{\Phi^{+i}}$&-&$\rfrac{\pi}{2}$&$\rfrac{\pi}{4}$&$\rfrac{\pi}{2}$&$\rfrac{\pi}{8}$\\
(vi)&$\ket{\Phi^{-i}}$&$\ket{VV}$&$\ket{\Phi^{+i}}$&$\rfrac{\pi}{4}$&$0$&$\rfrac{3\pi}{4}$&$\rfrac{3\pi}{8}$$^*$\\
(vii)&$\ket{\Phi^{-i}}$&$\ket{\Phi^{+1}}$&$\ket{\Phi^{+i}}$&$\rfrac{\pi}{4}$&$0$&$\rfrac{3\pi}{4}$&$\rfrac{\pi}{4}$$^*$\\
(viii)&$\ket{VV}$&$\ket{\Phi^{-1}}$&$\ket{HH}$&$\rfrac{\pi}{4}$&$0$&$\rfrac{\pi}{2}$&$0$\\
\hline\end{tabular}\end{center}
 \caption{\textbf{Experimental settings for pump laser preparation.} We show the target states and settings for 8 different pump preparations labelled (i)--(viii).  The corresponding target states in modes A--C are given; entangled states are expressed in the form  ${\ket{\Phi^\nu}=\frac{1}{\sqrt{2}}\left(\ket{HH}+\nu\ket{VV}\right)}$. The angles of the two $\alpha$-BBO crystals and the waveplates that follow are shown where a crystal angle of zero defines that a horizontally polarized beam is polarized along the slow axis. HWP-1 is always set to zero. Note that, in practice, settings (vi) and (vii) are subject to an additional phase due to wavelength-scale differences in the lengths of the two crystals; the states were set by rotating HWP-2 from the angle in the table until the two-photon measurements at D$_\mathrm{idler}$ and D$_\mathrm{in}$ matched the expected statistics. We indicated this experimental deviation from theory using the symbol $^*$ in the table.  This same additional phase also necessitates that the phase set by the tilted quarter-wave plate in the idler arm must be adjusted for setting (viii).}\label{PumpPrep}
\end{table}

\renewcommand{\arraystretch}{1.2}

\begin{table}[h!]
\arrayrulecolor{black}
  \begin{center}
       \begin{tabular}{|c|c|c|c|c|c|c|c|}
       \hline
       \multirow{2}{*}{Prep.}&\multirow{2}{*}{Det.}&Counts&\multicolumn{2}{c|}{Tangle}&\multicolumn{2}{c|}{Purity}&\multirow{2}{*}{Fidelity}\\ \cline{4-5}\cline{6-7}
       &&(cps)&meas.&theo.&meas.&theo&\\ \hline\hline

       \multirow{4}{*}{(i)}&in&$(45.36\pm0.10)\times10^3$&$0.8857 \substack{+0.0011 \\ -0.0011}$&1&$0.9435\pm0.0006$&1&$0.9605\pm0.0003$\\ \cline{2-8}\noalign{\vskip.2pt}
       &\bgd A&\bgd $0.67\pm0.06$&\bgd $0.0000 \substack{+0.0009 \\ -0.0000}$&\bgd 0&\bgd$0.349\pm0.018$&\bgd $\rfrac{1}{4}$&\bgd\\ \cline{2-8}\noalign{\vskip.2pt}
       &\bgd B&\bgd $0.42\pm0.05$&\bgd $0.000 \substack{+0.012 \\ -0.000}$&\bgd 0&\bgd$0.35\pm0.03$&\bgd $\rfrac{1}{4}$&\bgd\\ \cline{2-8}\noalign{\vskip.2pt}
       &C&$14.5\pm0.3$&$0.836 \substack{+0.015 \\ -0.014}$&1&$0.919\pm0.008$&1&$0.929\pm0.004$\\ \hline\hline

        \multirow{4}{*}{(ii)}&in&$(44.56\pm0.09)\times10^3$&$0.877\substack{+0.0010 \\ -0.0009}$&1&$0.9394\pm0.0005$&1&$0.9669\pm0.0003$\\ \cline{2-8}\noalign{\vskip.2pt}
       &\bgd A&\bgd $0.64\pm0.06$&\bgd $0.0000\substack{+0.0019 \\ -0.0000}$&\bgd 0&\bgd$0.35\pm0.02$&\bgd $\rfrac{1}{4}$&\bgd\\ \cline{2-8}
       &B&$13.9\pm0.3$&$0.828\substack{+0.011 \\ -0.011}$&1&$0.913\pm0.006$&1&$0.951\pm0.003$\\ \cline{2-8}\noalign{\vskip.2pt}
       &\bgd C&\bgd $0.28\pm0.04$&\bgd $0.000\substack{+0.006 \\ -0.000}$&\bgd 0&\bgd$0.30\pm0.03$&\bgd $\rfrac{1}{4}$&\bgd\\ \hline\hline

       \multirow{4}{*}{(iii)}&in&$(44.35\pm0.09)\times10^3$&$0.8807\substack{+0.0012 \\ -0.0012}$&1&$0.9409\pm0.0006$&1&$0.9581\pm0.0003$\\ \cline{2-8}\noalign{\vskip.2pt}
       &A&$12.5\pm0.3$&$0.737\substack{+0.019 \\ -0.020}$&1&$0.866\pm0.011$&1&$0.886\pm0.005$\\ \cline{2-8}\noalign{\vskip.2pt}
       &\bgd B&\bgd $0.49\pm0.06$&\bgd $0.000\substack{+0.006 \\ -0.000}$&\bgd 0&\bgd$0.34\pm0.03$&\bgd $\rfrac{1}{4}$&\bgd\\ \cline{2-8}\noalign{\vskip.2pt}
       &\bgd C&\bgd $0.39\pm0.05$&\bgd $0.000\substack{+0.005 \\ -0.000}$&\bgd 0&\bgd$0.38\pm0.03$&\bgd $\rfrac{1}{4}$&\bgd\\ \hline\hline

       \multirow{4}{*}{(iv)}&in&$(42.22\pm0.09)\times10^3$&$0.00005\substack{+0.00002 \\ -0.00002}$&0&$0.4857\pm0.0003$&$\rfrac{1}{2}$&$0.9794\pm0.0003$\\ \cline{2-8}\noalign{\vskip.2pt}
       &A&$7.26\pm0.14$&$0.00000\substack{+0.00004 \\ -0.00000}$&0&$0.854\pm0.006$&1&$0.921\pm0.003$\\ \cline{2-8}
       &B&$7.60\pm0.15$&$0.00000\substack{+0.00008 \\ -0.00000}$&0&$0.912\pm0.005$&1&$0.951\pm0.002$\\ \cline{2-8}\noalign{\vskip.2pt}
       &\bgd C&\bgd $0.32\pm0.03$&\bgd $0\substack{+0\\ -0}$&\bgd 0&\bgd$0.282\pm0.017$&\bgd $\rfrac{1}{4}$&\bgd\\ \hline\hline

       \multirow{4}{*}{(v)}&in&$(43.63\pm0.09)\times10^3$&$0.0001\substack{+0.0004 \\ -0.0003}$&0&$0.4843\pm0.0003$&$\rfrac{1}{2}$&$0.9812\pm0.0003$\\ \cline{2-8}\noalign{\vskip.2pt}
       &A&$7.77\pm0.15$&$0.658\substack{+0.014 \\ -0.014}$&0&$0.824\pm0.008$&1&$0.889\pm0.004$\\ \cline{2-8}
       &B&$6.42\pm0.13$&$0.769\substack{+0.014 \\ -0.012}$&0&$0.883\pm0.007$&1&$0.935\pm0.004$\\ \cline{2-8}\noalign{\vskip.2pt}
       &\bgd C&\bgd $0.26\pm0.03$&\bgd $0.000\substack{+0.003 \\ -0.000}$&\bgd 0&\bgd$0.300\pm0.021$&\bgd $\rfrac{1}{4}$&\bgd\\ \hline\hline

       \multirow{4}{*}{(vi)}&in&$(43.61\pm0.09)\times10^3$&$0.00019\substack{+0.00005 \\ -0.00004}$&0&$0.6074\pm0.0005$&$\rfrac{5}{8}$&$0.9843\pm0.0002$\\ \cline{2-8}\noalign{\vskip.2pt}
       &A&$3.97\pm0.11$&$0.48\substack{+0.02 \\ -0.02}$&1&$0.728\pm0.012$&1&$0.813\pm0.008$\\ \cline{2-8}
       &B&$7.01\pm0.14$&$0.001\substack{+0.007 \\ -0.001}$&0&$0.933\pm0.004$&1&$0.961\pm0.002$\\ \cline{2-8}
       &C&$3.78\pm0.10$&$0.62\substack{+0.02 \\ -0.02}$&1&$0.798\pm0.011$&1&$0.869\pm0.007$\\  \hline\hline

       \multirow{4}{*}{(vii)}&in&$(43.41\pm0.09)\times10^3$&$0.2075\substack{+0.0012 \\ -0.0013}$&$\rfrac{1}{4}$&$0.5953\pm0.0007$&$\rfrac{5}{8}$&$0.9760\pm0.0003$\\ \cline{2-8}
       &A&$3.36\pm0.10$&$0.40\substack{+0.02 \\ -0.02}$&1&$0.676\pm0.013$&1&$0.773\pm0.009$\\ \cline{2-8}
       &B&$6.79\pm0.14$&$0.720\substack{+0.012 \\ -0.014}$&1&$0.857\pm0.007$&1&$0.915\pm0.004$\\ \cline{2-8}
       &C&$3.51\pm0.10$&$0.58\substack{+0.02 \\ -0.02}$&1&$0.779\pm0.012$&1&$0.861\pm0.007$\\ \hline\hline

       \multirow{4}{*}{(viii)}&in&$(42.97\pm0.09)\times10^3$&$0.2165\substack{+0.0013 \\ -0.0013}$&$\rfrac{1}{4}$&$0.6010\pm0.0007$&$\rfrac{5}{8}$&$0.9780\pm0.0003$\\ \cline{2-8}
       &A&$3.24\pm0.09$&$0.0000\substack{+0.0006 \\ -0.0000}$&0&$0.716\pm0.011$&1&$0.832\pm0.007$\\ \cline{2-8}
       &B&$6.00\pm0.13$&$0.739\substack{+0.017 \\ -0.016}$&1&$0.867\pm0.009$&1&$0.913\pm0.004$\\ \cline{2-8}
       &C&$3.23\pm0.09$&$0.0002\substack{+0.0013 \\ -0.0002}$&0&$0.782\pm0.010$&1&$0.872\pm0.005$\\ \hline\hline

       \bgd \multirow{3}{*}{bkgd}
       \bgd &\bgd A&\bgd$0.55\pm0.04$&\bgd $0\substack{+0 \\ -0}$&\bgd 0&\bgd $0.332\pm0.013$&\bgd $\rfrac{1}{4}$&\bgd \\ \cline{2-8}\noalign{\vskip.2pt}
       \bgd (bkgd)&\bgd B&\bgd $0.34\pm0.03$&\bgd $0.0000\substack{+0.0007 \\ -0.0000}$&\bgd 0&\bgd $0.320\pm0.017$&\bgd $\rfrac{1}{4}$&\bgd \\ \cline{2-8}\noalign{\vskip.2pt}
       \bgd &\bgd C&\bgd $0.40\pm0.03$&\bgd $0.0000\substack{+0.0005 \\ -0.0000}$&\bgd 0&\bgd $0.317\pm0.015$&\bgd $\rfrac{1}{4}$&\bgd \\ \hline
\end{tabular}
\end{center}
\caption{\textbf{Full experimental results.} The preparations correspond to the eight preparations in Table~S-\ref{PumpPrep}, and each measurement was taken in coincidence with D$_\mathrm{idler}$. Grayed-out rows correspond to background counts. Uncertainties in tangle, purity, and fidelity were calculated with the assumption of Poissonian errors and a Monte Carlo calculation.}\label{FullResults}
\end{table}

\end{document}